\documentclass[a4paper,
               biblatex,     % biblatex is used
               ]{jacow}
\usepackage{pdfpages ,multirow ,ragged2e}
\usepackage{caption}
\usepackage{subcaption}

\makeatletter%
	\ifboolexpr{bool{xetex}}
	 {\renewcommand{\Gin@extensions}{.pdf,%
	                    .png,.jpg,.bmp,.pict,.tif,.psd,.mac,.sga,.tga,.gif,%
	                    .eps,.ps,%
	                    }}{}
\makeatother

\ifboolexpr{bool{xetex} or bool{luatex}}
 {}                                      
 {\usepackage[utf8]{inputenc}} 

\usepackage[USenglish]{babel}

\ifboolexpr{bool{jacowbiblatex}}%
 {%
  \addbibresource{TUZN01.bib}
 }{}
\begin{document}

\title{High-dimensional maximum-entropy \\phase space tomography\thanks{This manuscript has been authored by UT-Battelle, LLC, under Contract No. DE-AC05-00OR22725 with the U.S. Department of Energy.
The United States Government retains, and the publisher, by accepting
the article for publication, acknowledges that the United States Government retains a non-exclusive, paid-up, irrevocable, world-wide license
to publish or reproduce the published form of this manuscript, or allow
others to do so, for United States Government purposes. The Department of Energy will provide public access to these results of federally
sponsored research in accordance with the DOE Public Access Plan
(http://energy.gov/downloads/doe-public-access-plan)}
}

\author{
A.~Hoover\thanks{hooveram@ornl.gov}, Oak Ridge National Laboratory, Oak Ridge, TN, USA
}
	
\maketitle

\begin{abstract}

Reconstructing 4D or 6D phase space distributions from 1D or 2D measurements is a challenging inverse problem encountered in particle accelerators. Entropy maximization is an established method to incorporate prior information in the reconstruction, but it is typically infeasible in high-dimensional spaces. In this paper, I review two recent approaches to high-dimensional entropy maximization. The first approach utilizes differentiable simulations and a class of generative models known as \textit{normalizing flows}, whereas the second approach employs the method of Lagrange multipliers and Markov Chain Monte Carlo (MCMC) sampling. My aim is to provide a short explanation of each method using a common notation. I conclude by mentioning several unsolved problems in phase space tomography.
\end{abstract}

\section{Introduction}

Phase space density measurements enable model-based prediction and control of charged particle beams in particle accelerators. This paper focuses on indirect measurements in which the density is \textit{inferred} from partial information in the form of low-dimensional projections. Such measurements are especially challenging when the phase space is high-dimensional \cite{jaster-merz_experimental_2025}. One challenge is to search the space of distribution functions. Simple models may not be able to capture the full complexity of the real distribution, while complicated representations may require too many parameters to fit to the data \cite{wolski_transverse_2022}. The second challenge is that the data do not identify a unique distribution; rather, they carve out a set of \textit{feasible} distributions, each of which is compatible with the data to within the desired precision.

Here, I focus on the method of maximum entropy (MaxEnt) to select a distribution from the feasible set. After describing the reconstruction problem and MaxEnt strategy, I review two recent approaches to high-dimensional MaxEnt. The first approach leverages advances in differentiable simulations and generative modeling to find approximate maximum-entropy distributions. The second approach is an extension of the MENT algorithm, which uses the method of Lagrange Multipliers to find an exact constrained maximum entropy distribution. My aim is to provide a short explanation of each method using a common notation, rather then to provide a comprehensive literature review. I conclude by mentioning several unsolved problems in phase space tomography, including the preservation of dynamic range, the inclusion of self-forces in the forward model, and robust uncertainty quantification (UQ). I do not discuss experimental design---a separate but equally important problem.\footnote{Implementations of the methods described here, along with example applications, are found in software repositories \cite{github_ment, github_gpsr} and data repositories \cite{zenodo_ment_nd, zenodo_ment_4d, zenodo_ment_flow} linked to the corresponding papers \cite{hoover_n-dimensional_2025, hoover_high-dimensional_2024, roussel_efficient_2024}.}

\section{Maximum-Entropy Methods for Phase Space Reconstruction}\label{sec:review}

\begin{figure}
    \centering
    \includegraphics[width=1.0\columnwidth]{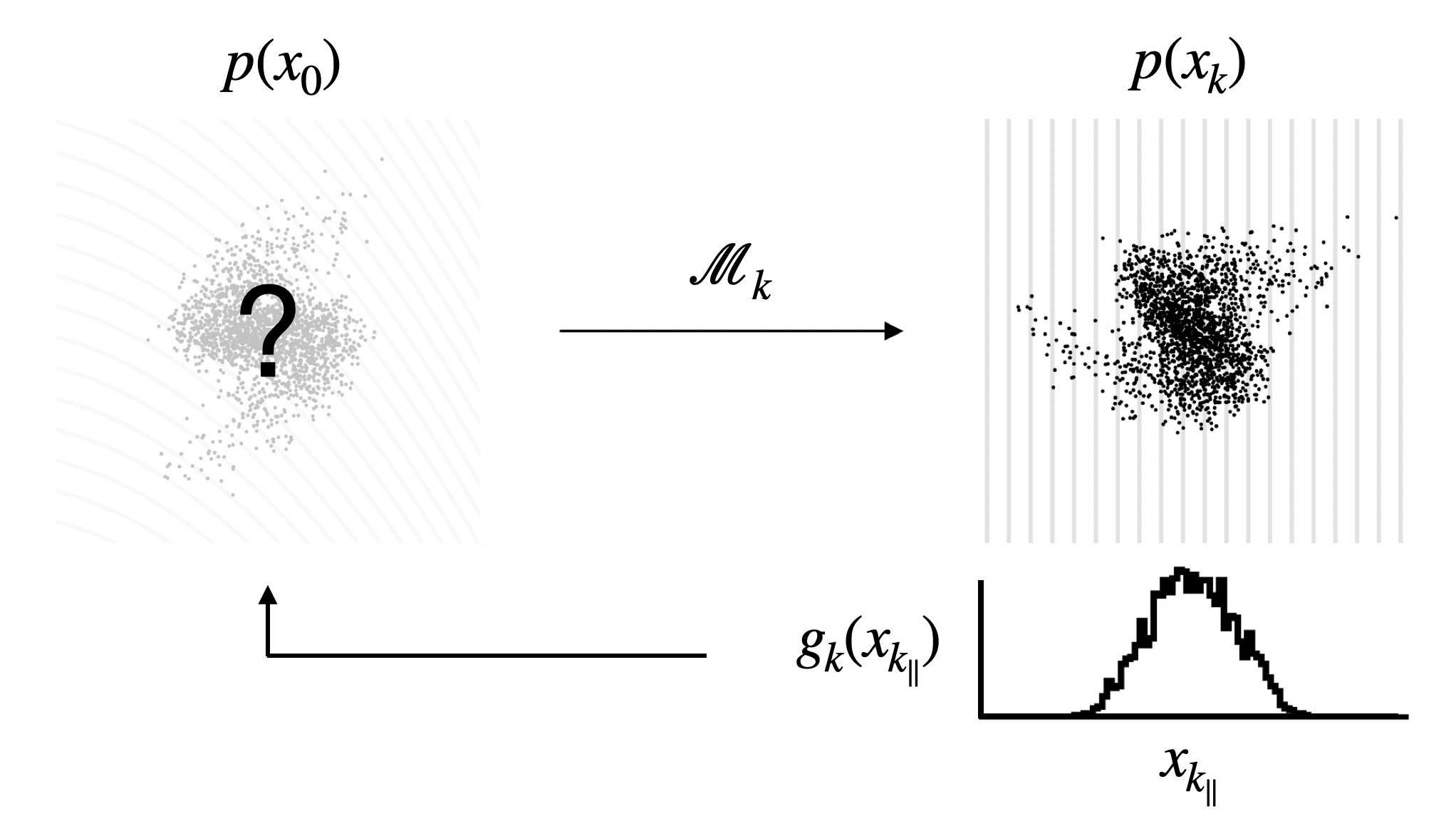}
    \caption{Illustration of the generic phase space reconstruction problem. We aim to recover a probability distribution from projections measured after a set of symplectic coordinate transformations. (Adapted from Ref.~\cite{hoover_high-dimensional_2024}.)}
    \label{fig:generic}
\end{figure}

We begin by describing the generic reconstruction problem. Let $p(x)$ be the probability density at location $x \in \mathbb{R}^N$ in phase space, where $N \in \{ 2, 4, 6 \}$ is the dimension of the phase space. Let $s$ be the distance along the accelerator, with the reconstruction location at $s = 0$. Let $\phi$ be a vector of accelerator parameters, such as quadrupole strengths, which determines the single particle dynamics between reconstruction and measurement location. For measurement indices $k = 1, \dots, K$, we select accelerator parameters $\phi_k$ and tracking distance $s_k$. These choices define a set of symplectic transformations from input coordinates $x \equiv x_0$ to output coordinates $x_k$:
\begin{equation} \label{eq:map}
    x_k = \mathcal{M}_k(x).
\end{equation}
We measure the projection of the transformed distribution onto a plane $x_{k_\parallel} \in \mathbb{R}^M$, where $1 < M < N$. Here, $x_{k_\parallel}$ identifies the observed coordinates on the measurement axis, while $x_{k_\perp}$ identifies the unobserved coordinates. With these definitions, the projection is written as
\begin{equation} \label{eq:proj}
\begin{aligned}
    g_k(x_{k_\parallel}) 
    &= \int p (\mathcal{M}_k^{-1}(x_k)) dx_{k_\perp}.
\end{aligned}
\end{equation}
This setup is illustrated in Fig.~\ref{fig:generic}.
We wish to recover the initial distribution $p(x)$. 

The data do not, in general, identify a unique distribution; the size of the solution space depends on the chosen coordinate transformations $\mathcal{M}_k$, the dimension of the phase space, and the dimension of the measurements. Rather than exploring the full solution space, let us instead search for a single distribution compatible with the data. Consider a \textit{prior} distribution $p_*(x)$ over the phase space coordinates; how should we update this prior in light of new information? We will perform the update by maximizing an entropy functional $S[p(x), p_*(s)]$ subject to the measurement constraints:
\begin{equation} \label{eq:opt}
    \max_{p(x)} S[p(x), p_*(x)] \quad \text{s.t.} \quad G_k[p(x)] = 0,
\end{equation}
where
\begin{equation} \label{eq:constraints}
    G_k[p(x)] = g_k(x_{k_\parallel}) - \tilde{g}_k(x_{k_\parallel}) = 0
\end{equation}
for measured projections $g_k$ and simulated projections $\tilde{g}_k$. The entropy functional is fixed by a set of axioms\footnote{See Refs.~\cite{caticha_maximum_2004, caticha_entropy_2021} for details.} which enforce the following principle: 
\textit{The chosen posterior distribution should coincide with the prior as closely as possible, and one should only update those aspects of one’s beliefs for which corrective new evidence has been supplied \cite{caticha_maximum_2004}.}
The resulting entropy is:
\begin{equation} \label{eq:entropy}
    S[p(x), p_*(x)] = - \int p(x) \log \left( \frac{p(x)}{p_*(x)} \right) dx,
\end{equation}
which lives in the range $[-\infty, 0]$ and reaches its maximum at the $p_*(x)$.

\begin{figure}
    \centering
    \includegraphics[width=0.70\columnwidth]{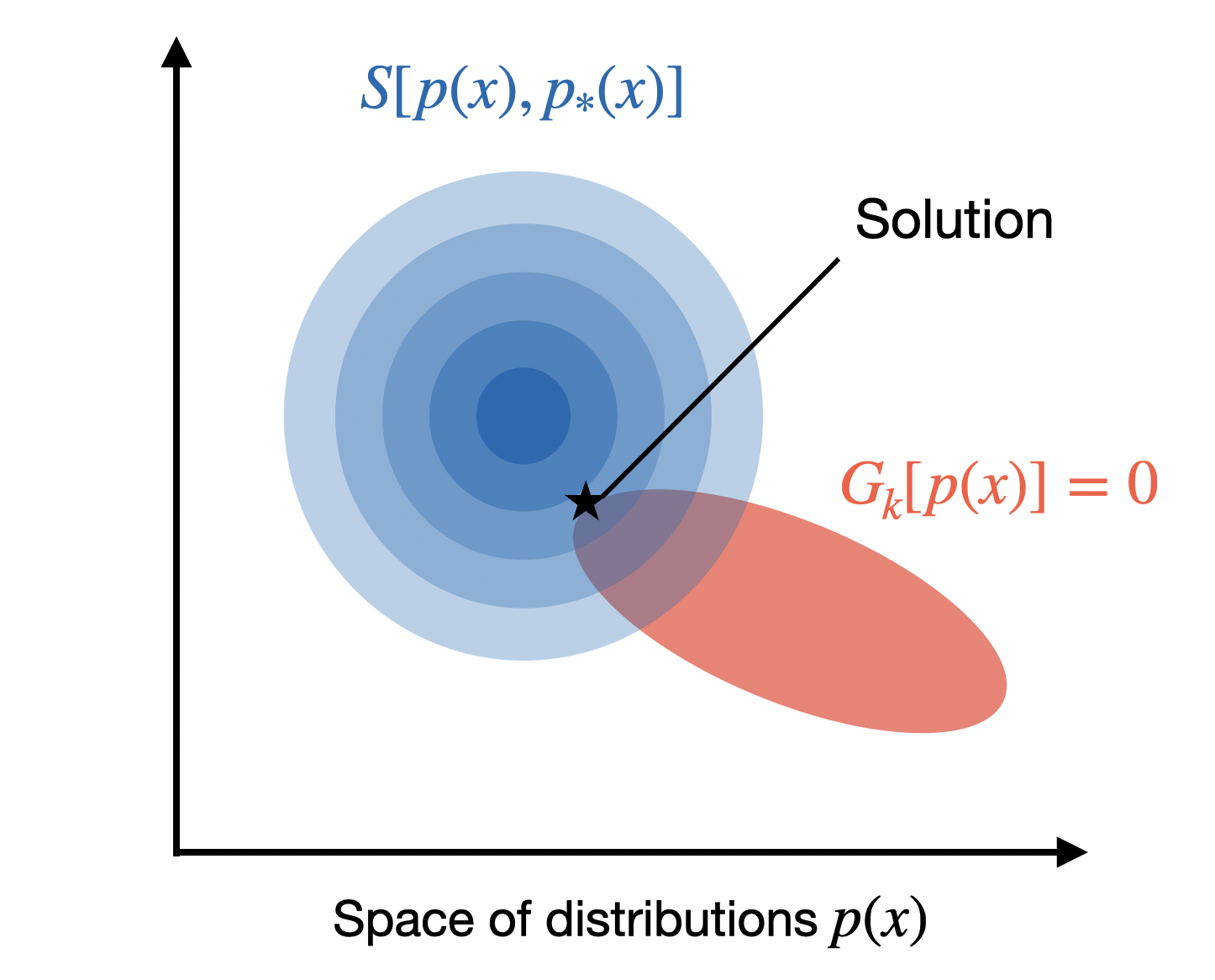}
    \caption{The method of maximum entropy (MaxEnt) selects the distribution of maximum relative entropy (blue) allowed by the measurement constraints (red). The solution is unique for the linear inverse problems described in this paper.}
    \label{fig:space}
\end{figure}

The optimization problem is illustrated in Fig.~\ref{fig:space}. 
The optimization is challenging both because of the difficulties in representing high-dimensional distributions and because of the hard constraints in Eq.~\eqref{eq:constraints}. The following sections describe two recent approaches to the problem.

\section{Generative modeling approaches} \label{sec:gen}

The first approach is based on the Generative Phase Space Reconstruction (GPSR) algorithm described in Ref.~\cite{roussel_efficient_2024}. Generative models typically sample particles from a probability distribution by first sampling a random variable $z \in \mathbb{R}^{N'}$ from a simple \textit{base distribution} $p(z)$, such as a Gaussian distribution, and then \textit{unnormalizing} the samples:
\begin{align}
    z &\sim p(z), \\
    x &= \mathcal{F}(z; \theta).
\end{align}
The unnormalized particles follow a distribution $p(x)$, even though the value of $p(x)$ may not be known. Thus, generative models provide easy-to-sample-from distributions of arbitrary complexity through the unnormalizing transformation $\mathcal{F}$. The unnormalizing transformation is typically implemented as an artificial neural network (NN) with parameters $\theta$, enabling efficient estimation of gradients with respect to $\theta$ using automatic differentiation (AD) techniques~\cite{baydin_automatic_2018}.

In GPSR, the transfer maps $\mathcal{M}_k$ in Eq.~\eqref{eq:map} are implemented using a differentiable physics simulation. This allows the propagation of gradients through the combined (unnormalization + lattice) transformation
\begin{equation}
    x_k = \mathcal{M}_k(\mathcal{F}(z; \theta)).
\end{equation}
The final step is the measurement operator, which is implemented by applying Kernel Density Estimation (KDE), a differentiable approximation to a histogram, to a batch of particles (projected onto the measurement axis). To simplify notation, introduce the function
\begin{equation} \label{eq:proj_func}
    \mathcal{P}_k(x) = x_{k_\parallel},
\end{equation}
which maps the phase space coordinates at $s=0$ onto the $k$th measurement axis. Then, given normalized particle coordinates $\{ z^{(i)} \}$, where $i$ is the particle index, the projected density is estimated as

\begin{equation}
\begin{aligned}
    g_k(x_{k_\parallel}) 
    &\approx 
    \text{KDE} 
    \left(
        \left \{
        \mathcal{P}_k(x^{(i)})
        \right \}
    \right)
    \\
    &\approx 
    \text{KDE} 
    \left(
        \left \{
        \mathcal{P}_k(\mathcal{F}(z^{(i)}; \theta))
        \right \}
    \right).
\end{aligned}
\end{equation}
GPSR has been demonstrated in 4D and 6D phase space tomography, both in simulated and experimental reconstructions \cite{roussel_phase_2023, roussel_efficient_2024}.

To incorporate entropy in GPSR, we restrict $\mathcal{F}$ to be a differentiable, invertible transformation called a \textit{diffeomorphism}. The change in probability density under such a transformation is given by
\begin{equation} \label{eq:flow_change_of_variables}
    \log{p({x})} = \log{p({z})} - \log{\left| {\det J_{\mathcal{F}}({z})} \right|},
\end{equation}
where $J_{\mathcal{F}} = dx/dz$ is the Jacobian matrix of the transformation. Generative models in which $\mathcal{F}: \mathbb{R}^N \rightarrow \mathbb{R}^N$ is a diffeomorphism are known as \textit{normalizing flows}, \textit{flows}, or \textit{flow-based models} \cite{papamakarios_normalizing_2021}. By tracking the change in density in each layer of the transformation, flows provide access to the exact probability density in addition to the exact samples from the model. Such models can therefore estimate high-dimensional integrals in the form of expectation values. We are interested in the entropy integral of Eq.~\eqref{eq:entropy}, which may be estimated from $n$ sampled particles $\{x^{(i)}\} \sim p(x)$ as follows:
\begin{equation} \label{eq:flow_entropy}
\begin{aligned}
    S[p({x}), p_*({x})]
    &= - \int p(x) \log(p(x) / p_*(x)) dx \\
    &= - \mathbb{E}_{p(x)}[\log(p(x) / p_*(x))] \\
    &\approx
    -\frac{1}{n} 
    \sum_{i=1}^{n} 
    \log ( p({x}^{(i)}) / p_*({x}^{(i)}) ) .
\end{aligned}
\end{equation}
The estimate is simply the mean of $\log(p(x)/p_*(x))$ over the batch of particles.

In Ref.~\cite{loaiza-ganem_maximum_2017}, Loaiza-Ganem, Gao, and Cunningham proposed the use 
of flows for entropy maximization. This technique was applied to the phase space reconstruction problem in Ref.~\cite{hoover_high-dimensional_2024}. It was found that an autoregressive flow architecture, with a masked neural network conditioner and rational-quadratic spline transformer, is a capable model with enough flexibility to represent complicated 6D distributions, but fast enough to train in a reasonable time. A simple penalty method was used with repeated minimization of the regularized loss function:
\begin{equation}
    L(\theta) = -S(\theta) + \mu \sum_k{D[g_k, \tilde{g}_k(\theta)]},
\end{equation}
where $g_k$ is the measured projection, $\tilde{g}_k$ is the simulated projection, and $D$ is a discrepancy function such as the mean squared error. The loss is minimized for a series of penalty parameters $\{\mu_1 < \mu_2 < \dots \}$, starting from $\mu_1 = 0$, until the reduction in prediction error is negligible. Sample generation is approximately 50\% slower than standard GPSR, and often around 5-10 penalty parameters are used, leading to typical training times of 5-20 minutes on a single GPU.

\begin{figure}
    \centering
    \includegraphics[width=\columnwidth]{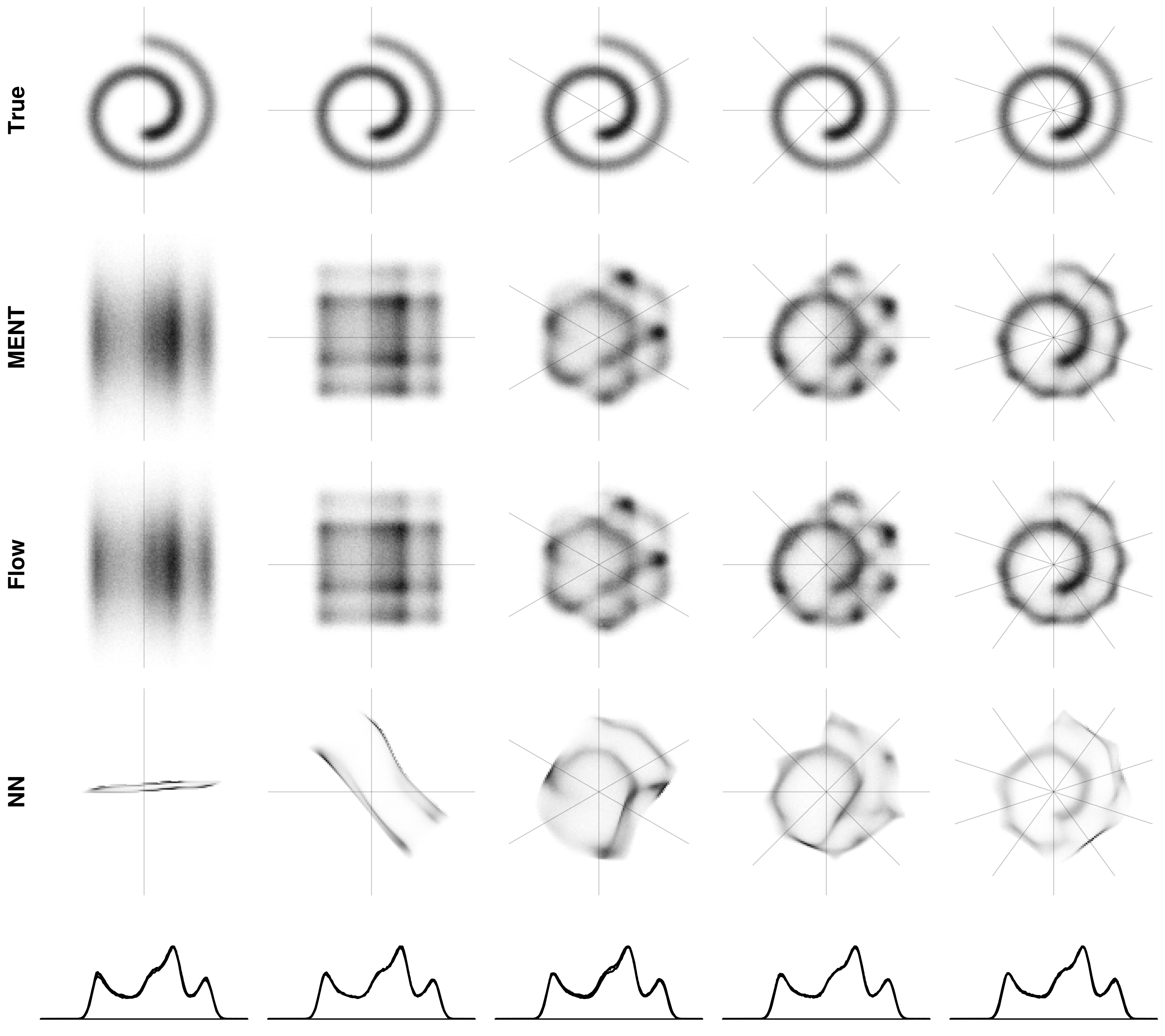}
    \caption{Entropy-regularized GPSR (MENT-Flow) can recover approximate maximum entropy distributions. The top row shows a ground-truth distribution, while the next three rows show MENT, MENT-Flow, and NN reconstructions with an increasing number of projections from left to right. The bottom row shows one of the projections in each case, with all four curves overlayed. (Adapted from Ref.~\cite{hoover_high-dimensional_2024}.)}
    \label{fig:swiss}
\end{figure}

The studies in Ref.~\cite{hoover_high-dimensional_2024} showed that the entropy penalty has a positive effect, encouraging smooth distributions and tending toward known 2D maximum-entropy solutions. Figure~\ref{fig:swiss} shows that the higher entropy distributions behave as expected for a Gaussian prior; for example, returning the product of the marginal distributions in the second column.
6D numerical experiments also show positive behavior compared to unregularized solutions, although comparison with exact maxent solutions is not possible.\footnote{Cross-validation is one approach to regularizing the behavior of NN fits without an explicit regularization function. No cross-validation was used in these tests.}

\section{Lagrange Multiplier Methods} \label{sec:ment}

The traditional approach to entropy maximization is the Lagrange Multiplier Method \cite{boyd_convex_2004}. This is the foundation of the MENT algorithm developed by Minerbo in 1971 \cite{minerbo_ment_1979}. We begin by constructing the Lagrangian $\Psi$:
\begin{equation}
    \Psi = S[p(x), p_*(x)] 
    + \sum_k
    \int
    G_k[p(x)] \lambda_k(x_{k_\parallel})
    dx_{k_\parallel}
\end{equation}
We then enforce zero change of $\Psi$ under variations in $\lambda_k$ and $p(x)$. The resulting expression for $p(x)$ is:
\begin{equation}
\begin{aligned}
    p(x) 
    &= p_*(x) \prod_k \exp(\lambda_k(x_{k_\parallel}))
    \\
    &= p_*(x) \prod_k h_k(x_{k_\parallel}).
\end{aligned}
\end{equation}
To make the dependence of $x_k$ on $x = x_0$ explicit, we make use of the projection function $\mathcal{P}_k$ in Eq.~\eqref{eq:proj_func}:
\begin{equation} \label{eq:ment_prob}
\boxed{
    p(x) = p_*(x) \prod_k h_k(\mathcal{P}_k(x)).
}
\end{equation}
Equation~\eqref{eq:ment_prob} gives the probability density at all points in phase space. Note that if we find Lagrange multiplier functions such that the distribution satisfies the constraints, we have found the \textit{exact} maximum-entropy distribution.

\begin{figure}
    \centering
    \includegraphics[width=\columnwidth]{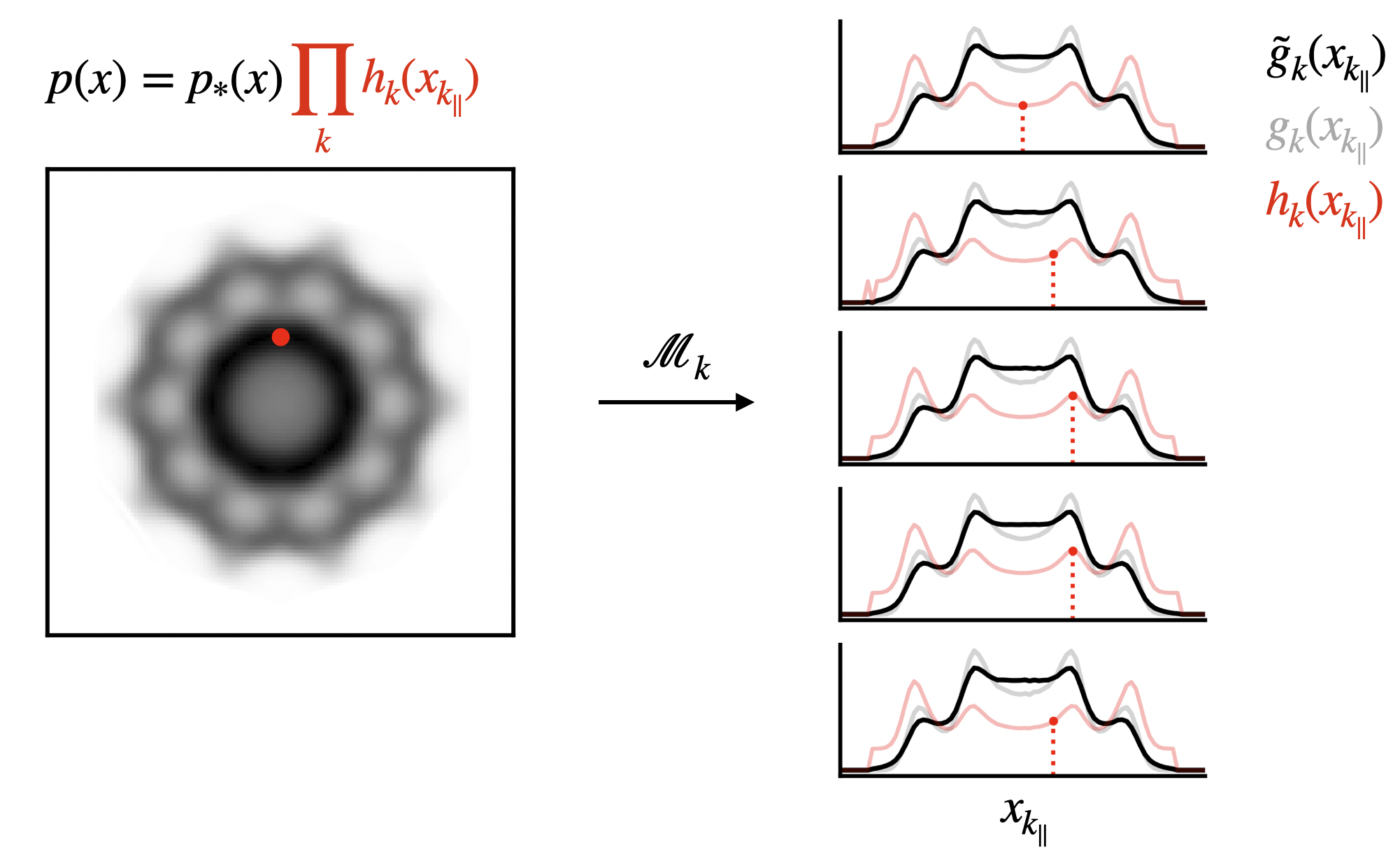}
    \caption{{Illustration of the MENT algorithm. The density, Eq.~\eqref{eq:ment_prob}, at a point is computed by mapping the point to each projection axis. The Lagrange multiplier functions $h_k$ (red) are multiplied together, and the resulting value is multiplied by the prior. Each iteration, Eq.~\eqref{eq:ment_gauss_seidel}, uses the simulated projections $\tilde{g}_k$ (black) and measured projections $g_k$ (grey) to update the Lagrange multipliers.}}
    \label{fig:ment}
\end{figure}

\begin{figure}
    \centering
    \includegraphics[width=\columnwidth]{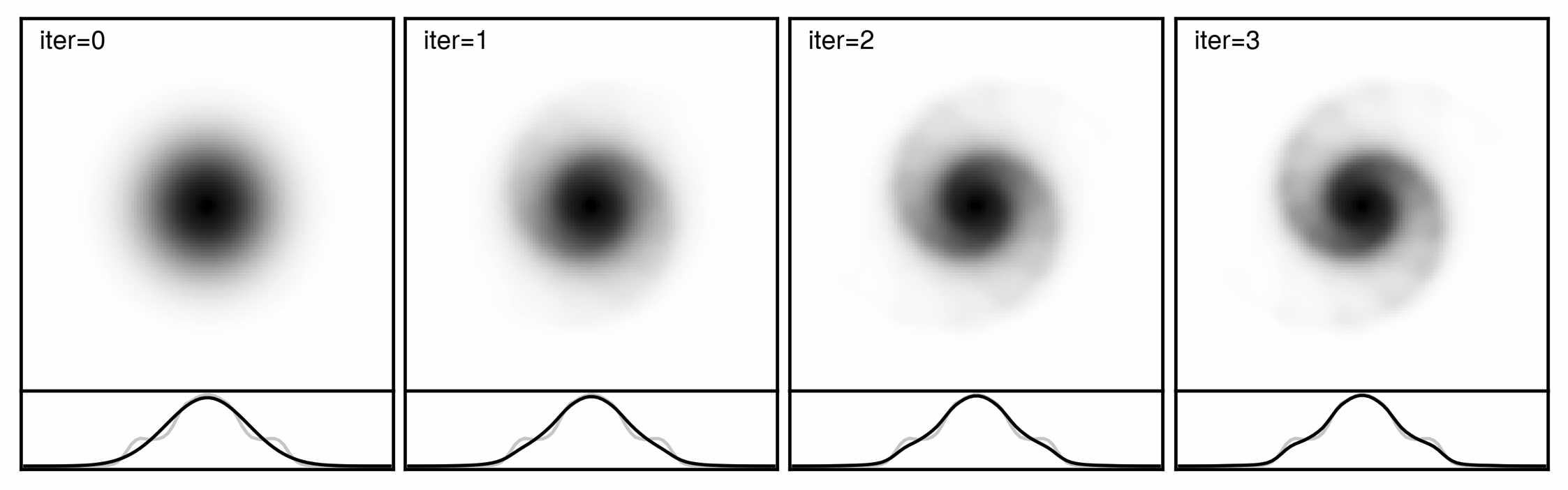}
    \caption{Maximum-entropy trajectory for small damping rate ($\omega$). Structure is gradually added to the Gaussian prior.}
    \label{fig:ment_path}
\end{figure}

Minerbo applied a nonlinear Gauss-Seidel method to solve for the Lagrange multiplier functions \cite{minerbo_ment_1979}. Plugging the distribution function, Eq.~\eqref{eq:ment_prob}, into the constraint equations, Eq.~\eqref{eq:constraints}, generates a simple algorithm that converges to the optimal distribution. The functions are initialized to constant values so that the distribution starts at the prior. Then the functions are updated as
\begin{equation} \label{eq:ment_gauss_seidel}
    h_k^{(i + 1)} =
    h_k^{(i    )}
    \left[
        (1 - \omega) + \omega
        \left(
            \frac{{g}_k}{\tilde{g}_k}
        \right)
    \right],
\end{equation}
for damping rate $0 < \omega \le 1$. Visual aids are provided in Figs.~\ref{fig:ment} and \ref{fig:ment_path} shows an example maximum-entropy trajectory for a small value of $\omega$. For typical values of $\omega$, MENT converges in 1-2 iterations.

\begin{figure*}
     \centering
     \begin{subfigure}[]{\columnwidth}
         \centering
         \includegraphics[width=0.95\columnwidth]{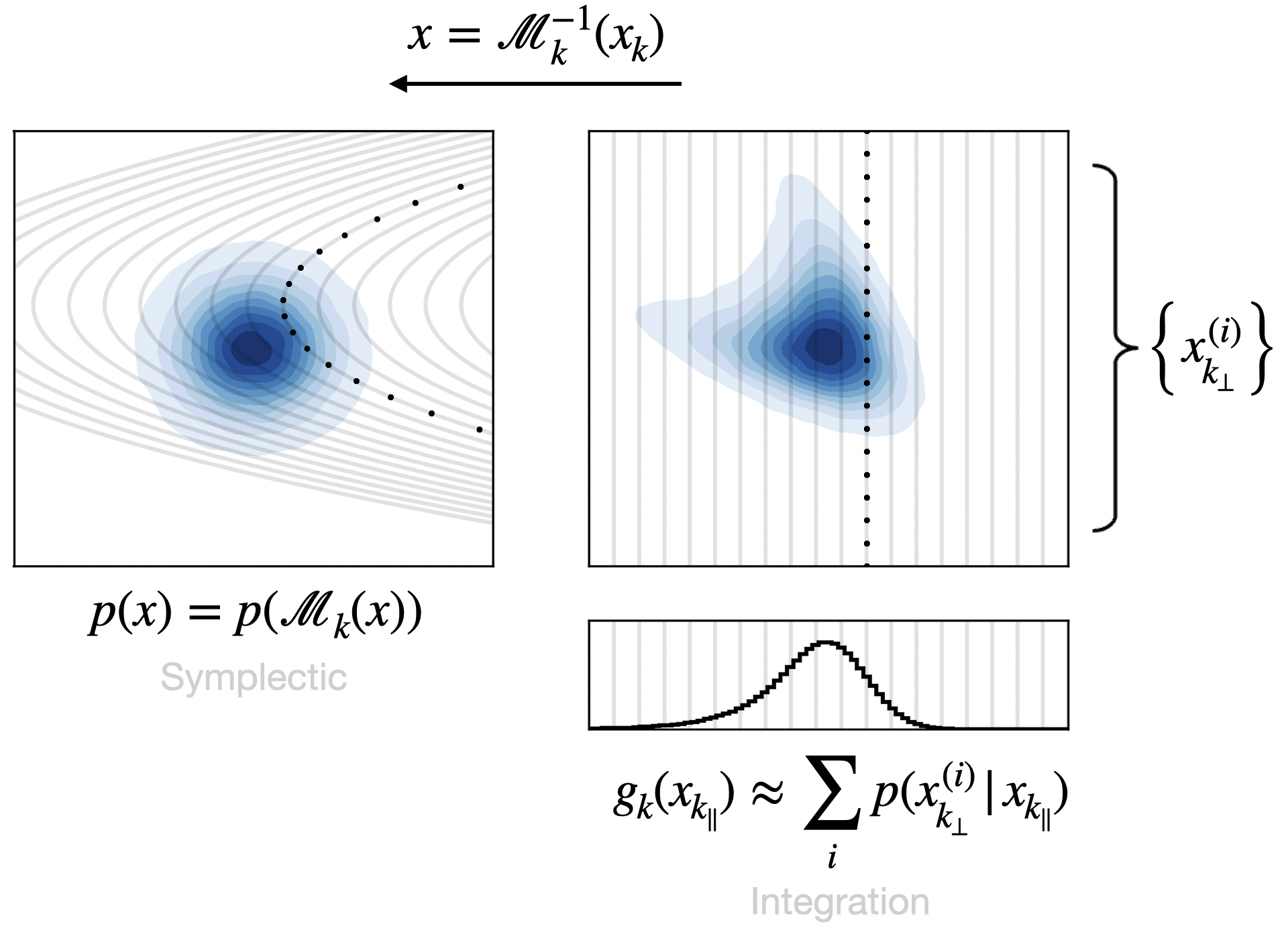}
         \caption{Numerical integration (reverse mode)}
         \label{fig:ment_diag_a}
     \end{subfigure}
     \hfill
     \begin{subfigure}[]{\columnwidth}
         \centering
         \includegraphics[width=0.95\columnwidth]{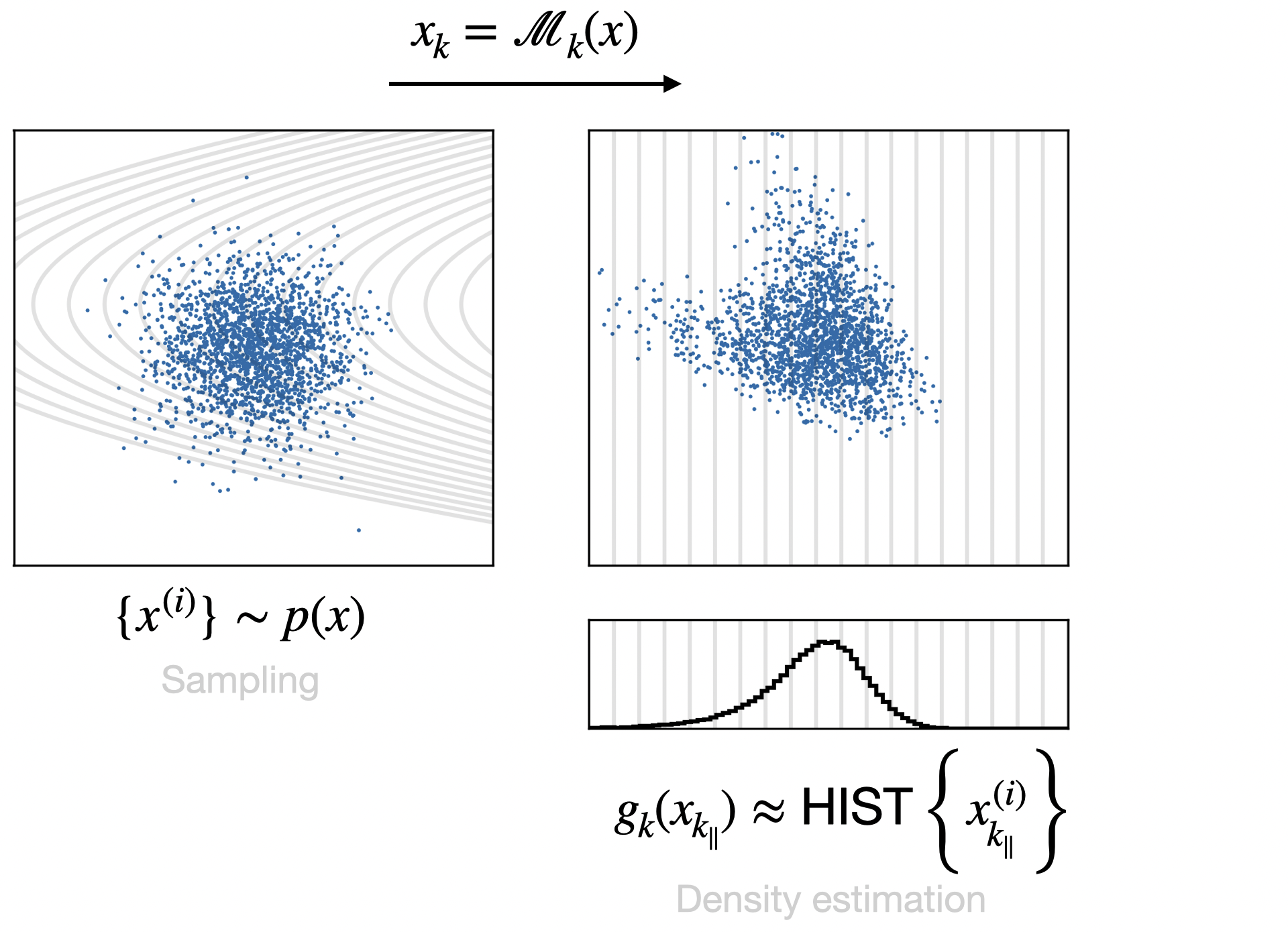}
         \caption{Particle sampling (forward mode)}
         \label{fig:ment_diag_b}
    \end{subfigure}
    \caption{Two ways to implement the Gauss-Seidel iterations in the MENT algorithm. (a) Numerical integration: integration points are mapped backward to the reconstruction point to evaluate the probability density, then summed over to compute the projection. (b) Particle sampling: particles are sampled from the distribution and mapped forward to the observation point, and the projection is estimated using a histogram.}
    \label{fig:ment_diag}
\end{figure*}

MENT was originally applied to a 4D reconstruction in 1981 \cite{sanders_beam_1979}. This high-dimensional application of MENT was revived in Ref.~\cite{wong_4d_2022} and continued in Ref.~\cite{hoover_four-dimensional_2024} at the Spallation Neutron Source (SNS) accelerator. Application to 6D tomography is more challenging. Each MENT iteration involves mapping the distribution function in Eq.~\eqref{eq:ment_prob} to its projections. Conventional implementations use numerical integration to compute the integrals as shown in Fig.~\ref{fig:ment_diag_a}. 
When the phase space is six-dimensional, the numerical integrals become four- or five-dimensional. The integrals are repeated for each point on the measurement axis; for example, at each pixel of a measured image. This numerical integration has a large computational cost making it infeasible for most 6D problems. For example, consider 15~projections of a 4D distribution onto a $50 \times 50$ measurement grid using a $50 \times 50$ integration grid. Each MENT iteration involves $15 \times 50 \times 50$ integrals. On a single CPU, each MENT iteration takes 60 seconds. When the distribution is 6D and the integration grid is $50 \times 50 \times 50 \times 50$, each MENT iteration takes \textbf{26 hours}.\footnote{\url{https://github.com/austin-hoover/ment/blob/main/examples/rec_nd_2d/train_nd_marg_proj_reverse.ipynb}.}

I propose to compute the projections by sampling particles from the distribution function, as illustrated in Fig.~\ref{fig:ment_diag_b}. In Ref.~\cite{hoover_n-dimensional_2025}, this approach was demonstrated with the Metropolis-Hastings (MH) algorithm, a simple variant of Markov Chain Monte Carlo (MCMC) sampling \cite{gelman_bayesian_1995}. MH evolves particles through the phase space via a random walk such that, in the long run, the chain of particles converges to the target distribution. In Ref.~\cite{hoover_n-dimensional_2025}, MH was used to demonstrate 6D MENT convergence on both a multimodal synthetic distribution as well as on experimental data, shared in Ref.~\cite{roussel_efficient_2024}. The runtime for the experimental 6D reconstruction, which involved 20~projections, was 7 minutes per iteration, converging in a few iterations.

\section{Unsolved problems} \label{sec:unsolved}

\subsection{Uncertainty Quantification}

The algorithms described here offer paths to maximum-entropy inference of high-dimensional phase space distributions. But even if these algorithms work perfectly, they do not provide a complete solution to the reconstruction problem. A complete solution would assign a probability to \textit{all possible distributions}, then update these probabilities in response to data. We consider this a posterior distribution over the space of probability density functions. The maximum of the posterior, or MAP estimate, represents the most probable distribution (if there is a single maximum) and is thus a sensible choice if a single answer is desired. However, if the posterior is not sharply peaked, then there will be large uncertainty in this estimate.

In practice, we usually introduce a set of parameters $\theta$ to represent the distribution $p(x)$, writing either $p(x | \theta)$ or $p_\theta(x)$. We could also represent the measured projections using vectors $\eta_k$. Then, given a prior $P_*(\theta)$, we write the posterior probability of $\theta$:
\begin{equation}
    P(\theta) \propto P_*(\theta) \prod_k P(\eta_k | \theta).
\end{equation}
The likelihood $\prod_k P(\eta_k | \theta)$ represents the probability of the data, given $\theta$. With an error-free forward model and noiseless measurements, the likelihood is constant for distributions allowed by the constraints and zero otherwise. If we include random errors in the forward model, the likelihood will vary smoothly.

The prior $P_*(\theta)$ assigns a probability to each distribution logically prior to observing the data. Caticha \cite{caticha_maximum_2004} notes that the parameterized family of distributions $p(x | \theta)$ forms a \textit{statistical manifold} with a unique metric known as the \textit{Fisher information metric}. Caticha uses this metric to extend the maximum entropy technique beyond a point estimate and generate a full posterior over $\theta$. It is unclear whether other choices of prior are justified based on simulated data or other assumptions.

If we can agree on a prior and likelihood, the next step is to characterize the posterior. A first step would be to quantify the posterior's width at its maximum, which provides a notion of uncertainty for the MAP estimate. This could be possible using a Gaussian approximation \cite{krause_probabilistic_2025}. A next step would be to sample different $\theta$ vectors from the posterior. The samples represent an ensemble of distributions from which various statistics and visualizations could be produced. Sampling from the posterior seems to be a straightforward but possibly very high-dimensional sampling problem that could be attacked with a variety of modern algorithms.

Finally, it is important to include uncertainty in the accelerator lattice parameters $\phi$, such as quadrupole strengths, which may not be known precisely \cite{wolski_accelerator_2024}. In the Bayesian framework, this would require characterizing the joint posterior distribution $P(\theta, \phi$) over both the lattice and beam parameters.

\subsection{Dynamic Range}

Low-density regions of phase space are important for predicting beam loss in high-power hadron accelerators \cite{aleksandrov_understanding_2020}. Measurements sensitive to these low-density regions are considered \textit{high-dynamic-range} (HDR), where the dynamic range is defined as the ratio of maximum to minimum measured density. Thus far, tomographic reconstructions have focused on low-dynamic-range (LDR) projections. GPSR is expected to struggle with HDR data, but MENT appears to be well-suited based on 2D tests. The limit on dynamic range in 4D/6D reconstructions is unknown.

\subsection{Collective Effects}

The inclusion of collective effects in the forward model dramatically complicates the reconstruction problem. In this case, the coordinate transformations in Eq.~\eqref{eq:map} depend on the initial distribution:
\begin{equation}
    x_k = \mathcal{M}_k(x, p(x)).
\end{equation}
The resulting inverse problem is nonlinear, and most standard algorithms do not apply. On the other hand, it seems plausible that collective effects could place very strong constraints on the initial distribution, reducing the number of measurements required for high-dimensional reconstructions. The differentiable space charge model proposed in Ref.~\cite{qiang_differntiable_2023} could be used in a GPSR forward model, although it is unknown if this is feasible. It is unclear how to incorporate space charge into MENT.

\section{Summary}

This article has reviewed two algorithms for maximum-entropy inference of phase space distributions from projection data. 

GPSR is an efficient method to search the space of high-dimensional probability distributions if differentiable simulations are available. The entropic regularization discussed here could be improved by exploring new flow architectures to increase sampling speed, increase model flexibility, and decrease model size. In particular, it may be possible to use non-invertible unnormalizing transformations and directly compute the Jacobian matrix using automatic differentiation. The penalty method used to train the model could also be substituted for a more sophisticated constrained optimization algorithm with an automated stopping condition \cite{schechtman_orthogonal_2023}.

MENT is a surprisingly simple algorithm to fit high-dimensional distributions to measured projections. Its primary advantages are (i) exact constrained maximization, (ii) fast convergence, (iii) minimal storage requirements, (iv) no restrictions on the forward model, (v) the ability to represent every possible distribution function, and (vi) preservation of large dynamic range. In 2D and 4D tomography, MENT is an established method. The 6D (ND) version of the algorithm works but requires hand-tuning of the MCMC sampling algorithm. Future work will examine modern adaptive samplers \cite{bou-rabee_nurs_2025} to more efficiently explore the phase space and more systematically tune the sampling parameters. It will also be important to parallelize the algorithm, which could provide significant speedups.

\printbibliography

\end{document}